\documentclass[conference]{IEEEtran}
\IEEEoverridecommandlockouts
\usepackage{cite}
\usepackage{amsmath,amssymb,amsfonts}
\usepackage{algorithmic}
\usepackage{graphicx}
\usepackage{textcomp}
\usepackage{xcolor}
\usepackage[left=1.57cm,right=1.57cm,top=0.95cm,bottom=2.54cm]{geometry}

\def\BibTeX{{\rm B\kern-.05em{\sc i\kern-.025em b}\kern-.08em
    T\kern-.1667em\lower.7ex\hbox{E}\kern-.125emX}}
\begin{document}

\title{Is FIDO2 Passwordless Authentication \\ a Hype or for Real?: A Position Paper 
\thanks{This research is funded by TUBITAK (The Scientific and Technological Research Council of Turkey) under the grants No: 3211046 and No:3200184.}
}

\author{\IEEEauthorblockN{Kemal Bicakci}
\IEEEauthorblockA{\textit{Informatics Institute} \\
\textit{Istanbul Technical University}\\
Istanbul, Turkey \\
0000-0002-2378-8027}
\and
\IEEEauthorblockN{Yusuf Uzunay}
\IEEEauthorblockA{\textit{Securify Information Tech. and Security} \\
\textit{Training Consulting Inc.}\\
Ankara, Turkey \\
0000-0001-8768-6620}
}

\maketitle

\begin{abstract}
Operating system and browser support that comes with the FIDO2 standard and the biometric user verification options increasingly available on smart phones has excited everyone, especially big tech companies, about the passwordless future. Does a dream come true, are we finally totally getting rid of passwords? In this position paper, we argue that although passwordless authentication may be preferable in certain situations, it will be still not possible to eliminate passwords on the web in the foreseeable future. We defend our position with five main reasons, supported either by the results from the recent literature or by our own technical and business experience. We believe our discussion could also serve as a research agenda comprising promising future work directions on (passwordless) user authentication.
\end{abstract}


\section{Introduction}

We have been discussing replacing passwords at least since 2004 when then-chairman of Microsoft Bill Gates predicted their death. Although almost a consensus has been established about the numerous usability and security problems of passwords, despite many attempts, up until now, passwords remain a de facto standard for web authentication. Leaving a more elaborate technical analysis of the status quo to understand why replacing passwords is not easy to a seminal paper in the field written by Bonneau et al. \cite{bonneau}, we present a shorter discussion specifically targeting the FIDO2 (Fast IDentity Online) standard here. We can summarize what this recent standard provides that makes many of us more optimistic about the passwordless future as follows:

Broadly speaking, we know that there are three main categories of solution for user authentication, based on either (i) something you know, or (ii) something you have, or (iii) something you are or you do. Hence, obvious choices to replace the first one in this list, which is the passwords, should be based on either second or third or both of these categories:

1) Something you have: This comes either in the form of dedicated piece of hardware such as a smartcard or USB stick or personal device like a smart phone. The main reason why we usually could not depend solely on something you have factor is the risk of losing it \cite{lyastani}. Somebody, who obtains physical possession of an authenticator device, could easily impersonate its owner. This risk is mitigated by requiring a PIN code as a second factor. However the PIN undermines the benefit of allowing passwordless authentication \cite{farke}.

2) Something you are or you do: This comes either in the form of physical or behavioral biometrics. The latter has high error rates hence could not be deployed as a sole method \cite{killourhy}. The error rates of physical biometrics such as fingerprint or face recognition are usually considered as low enough. Their real problem when used directly for web (remote) authentication is the security and privacy risks due to server side storage of biometric data. Unlike passswords, once stolen, you cannot change your biometric information. A more acceptable application scenario is the use for local authentication e.g., to unlock a mobile phone where biometric data is not shared by any other party.

3) Combination of (1) and (2): The ability to control access to an authenticator device with biometrics has the potential to solve problems of both (1) and (2). A stolen device is useless by itself if we could assume biometrics data is stored securely on it and not shared. However we argue that this option has also drawbacks:

1) Integrating biometrics with an authenticator device is not always economically feasible. 


2) Most modern mobile phones already have support for biometrics hence the above cost argument becomes less of an issue. On the other hand, this support is less common in laptops, desktops and others\footnote{Even when the support is available, there has to be still standard interfaces available to browsers to facilitate its use.}.

After several earlier attempts by FIDO Alliance, its latest FIDO2 standard is the industry's answer to these problems: why not use the smart phone\footnote{Although there are other FIDO2 use cases e.g., with a plugged USB device or with a platform authenticator which includes TPM (Trusted Platform Module), in this paper we restrict our discussion to the use of smart phones as FIDO2 authenticators since we consider it as a more usable and deployable option in general.} as a roaming authenticator while using our laptop? 

What makes FIDO2 special is its standard interfaces. CTAP2 (Client to Authenticator Protocol) is used for the communication between the browser and the authenticator and WebAuthn (Web Authentication) protocol defines a web browser API to enable web sites to offer passwordless authentication. Major web browsers (Google Chrome, Mozilla Firefox, Microsoft Edge, Apple Safari) and operating systems (Windows, MacOS, Linux, Android) have a support for FIDO2 \cite{news}. FIDO2 specifications are available online \cite{webauthn} \cite{ctap2}. 

If we could assume the user has already completed the registration phase, the user experience of web authentication with FIDO2 could be as easy as the following three steps:

1) The user attempts to authenticate on a web site.

2) She will receive a push notification on her smart phone.

3) She can accept the notification by tapping after unlocking her device with biometrics e.g., with her fingerprint \footnote{According to latest FIDO2 WebAuthn specification \cite{webauthn}, the web site could choose to enforce user verification on the smart phone but it is up to the user to choose either a PIN or biometrics for this verification.}.

We summarize the advantages of the described FIDO2 passwordless authentication over password based authentication as follows:

1) FIDO2 protocols use standard public key cryptography techniques for user authentication. There is no need for the user to enter a password and for the server to store any data derived from passwords. Elimination of passwords eliminates all password attacks including offline brute force and online guessing.

2) A security advantage of FIDO2 worth considering separately is the real-time phishing protection. While standard multi-factor authentication could eliminate most password attacks including traditional forms of phishing, it is vulnerable to a form of social engineering attack called real-time phishing. Here, the attacker relays second-factor one-time tokens from the victim user visiting the phishing site to the legitimate web site in real-time or (he tricks or simply expects her to accept the notification on her phone without reading since the victim most probably has already been conditioned to tap on the accept button). In FIDO2's challenge-response mechanism, the URL of the visited web site is included in the challenge by the browser. Thus, relaying the response becomes useless hence real-time phishing is prevented \cite{ulqinaku}.


3) Besides security advantages briefly discussed above, FIDO2 standard arguably offers better usability since passwordless authentication avoids all the frustration that comes with passwords. 

4) Last but not the least, FIDO2 could help organizations cut costs. For instance elimination of passwords means no helpdesk calls for password resets\footnote{But there might be other reasons to call the helpdesk, as we will elaborate in section 3.}.

We argue that even though at least some of the aforementioned advantages are real and significant, change to FIDO2 passwordless authentication will still be not easy and straightforward. The following sections defend our position with various reasons and their supporting details.



\section{Workforce Identity vs. Customer Identity}

Unlike the industry where the distinction is clear and well-known, many previous academic work did not acknowledge the differences between workforce identity and customer identity in the context of authentication problem. We start our discussion by emphasizing this major point.

Workforce identity management systems are designed to manage access of employees. Customer identity is about making business with customers.  To make a long story short, generally speaking, security is more important than usability in workforce use cases. On the other hand, increasing engagement and therefore usability is essential while dealing with customers. Therefore, unlike workforce use cases where strict and final decisions could be given for risk minimization, passwordless authentication could only be one of the choices when offered to customers. The final decision should be left to customers\footnote{We observe that classical types of 2FA is also not enforced in almost all commercial websites. The only exception is regulated industries like banking.}. Otherwise, there is a significant risk of losing impatient customers to competitors (and unfortunately we know that customers are impatient most of the time). As a result, customers accustomed to passwords either choose to change their login routine or not while interacting with your web site.

This opening passage raises two important questions, one for each identified use case:

1) If it is kept optional, is it reasonable to expect most customers make a transition to passwordless authentication?

2) Do decision makers in organizations prefer passwordless authentication for their workforces? 

Our answer to the first question is a certain no. The principle of path-of-least resistance states that given a set of alternative paths, users choose the path with the least amount of effort required (thus, this path should be the secure one \cite{yee}).

We argue that path-of-least resistance for most users is not to take the hassle of change. The discomfort of passwords usually does not reach the level demanding an effortful change. By time, users have already developed many coping strategies for password management \cite{bonneau}. Besides, the earlier work showed the transition to passwordless authentication is far from being painless.

In a usability study conducted by Oogami et al., participants faced significant impediments in the setup process for passwordless authentication \cite{oogami}. Only 3 out of 10 users could complete authenticator registration using Android Phones with fingerprint.  In another study (although conducted using Yubikey hardware tokens not smart phones, we still think their results could provide a general idea about problems in the setup phase of passwordless authentication), only a third of the participants successfully configured the YubiKey to work with their Facebook account \cite{reynolds}. The setup phase of FIDO is a high inconvenience for users due to lacking instructions and guidance. User interface design choices also contribute to the problem \cite{ciolino}.

FIDO2 is a new standard. Obviously, some of the technical matters in the setup process could be solved by time. However we still think these results from recent literature hint us that setup for passwordless authentication is not frictionless. (Secure or not) passwords remain as the path-of-least resistance in the foreseeable future.
\\

\fbox{\begin{minipage}{24em}
REASON-1: Passwords will remain path-of-least resistance unless passwordless authentication is enforced (not an option for customer identity).
\end{minipage}}
\\

Returning back to workforce identity use case where a stronger alternative to passwords could certainly be enforced for security reasons, the decision to be made is whether to eliminate the passwords all together or use an additional security layer beside passwords (referred earlier as FIDO U2F in FIDO Alliance jargon).

The proponents of FIDO2 passwordless authentication could argue that passwordless authentication implemented with local biometric verification is secure against many prominent threats and there is no need for passwords. We agree that modern features of smart phones such as FaceID and the protection of cryptographic keys using technologies such as Secure Enclave Processor \cite{gulati} significantly raises the bar against existing threats. On the other hand, history teaches us that no security system remains impenetrable\footnote{Although not directly targeting security properties, we want to mention a recent work \cite{kepkowski} which presents a timing attack on FIDO2 protocols that allows attackers to link user accounts, which is a serious privacy concern.}.  There is some sort of risk involved and despite how small that risk is, security decision makers may choose to keep user passwords for risk mitigation. Passwords could be considered one extra layer of protection by many. Yet some others could see it an illusion of stronger security though inflicting needless pain and cost. Time will tell which side prevails but we can see that two critical aspects will play a role: 

1) We learn from the past that previously established practices, later shown to have only marginal security gains but significant usability problems, cannot be terminated easily. This could be attributed to the risk averse decision makers. For instance, PCI-DSS security standards have not kept up and continue to promote password expiration policies although many recent security guidelines advice against them.

2) While the inconvenience of dealing with many passwords which also brings extra cost may be rejected easier in favor of passwordless authentication, managing a single (master) password could be considered as acceptable for SSO (single sign on) systems. Many enterprises have already completed successful SSO projects which allow users (employees) to authenticate only once no matter how many applications are being accessed during a session. 
\\

\fbox{\begin{minipage}{24em}
REASON-2: Three-factor authentication with passwords will be regarded as more secure (and preferable for SSO logins) than passwordless authentication. 
\end{minipage}}

\section{Use and Abuse of Recovery Mechanisms}

In many user studies on FIDO solutions, a predominant concern among the participants was the loss of the authenticator device, which they feared would stop them accessing their accounts \cite{lyastani}. Recommendation of FIDO Alliance to address this issue is adding additional authenticators \cite{recovery} used for account recovery, which we believe is neither practical nor cost-effective. 

In practice, we see that many web sites deploy alternative 2FA schemes as account recovery mechanisms, instead. In fact, as of 2021, all FIDO2-supporting sites in Alexa’s top 100 allow choosing alternatives \cite{ulqinaku}. Unfortunately, social engineering attacks in which an attacker downgrades users to an authentication mechanism vulnerable to real-time phishing attacks are shown to be quite effective. In a carefully design user study, Ulqinaku et al. showed that 55\% of participants fell for real-time phishing, and another 35\% would potentially be susceptible against a FIDO-downgrade attack \cite{ulqinaku}.

Kunke et al. provided a comprehensive evaluation of 12 account recovery mechanisms for FIDO2 passwordless authentication \cite{kunke}. They also showed that the currently used methods have many drawbacks. They reported that some of these methods even rely on passwords, which simply makes the whole idea of passwordless authentication nonsense. 

Currently, there is a lack of adequate standardized recovery method for FIDO2 passwordless authentication \cite{kunke}. In an enterprise setting, we would expect a lot of helpdesk calls; not for password reset but for passwordless reset instead.
\\

\fbox{\begin{minipage}{24em}
REASON-3: There is no convenient and secure recovery method for FIDO2 passwordless authentication, yet.
\end{minipage}}

\section{Shared Accounts and Access Delegation}

An overlooked aspect of the password replacement problem is shared accounts and access delegation.

An easy way to share accounts and delegate access is password sharing, which is a frowned-upon but frequently used method. It was reported that nearly everyone (95\% of respondents) share passwords despite the fact that many of them (73\% of survey respondents) agree it is risky \cite{lastpass}.

In FIDO2 passwordless authentication with a smart phone, to share an account with a person, the user has to give her phone (unless more than one device is registered). This is an inconvenience (might be even more annoying than giving a piece of dedicated hardware) and sometimes impossible (when not physically be in the same place) \cite{fatima}. Recall that in FIDO2, the signature generated by the authenticator has to be transferred to the web site over the browser holding the active session. 

Not all multi-factor solutions require transferring possession of authenticator devices to share accounts. Consider a traditional form of 2FA implemented with push notification and assume that the password is already shared. The only extra thing for the account owner is to accept the notification remotely. One can even claim that this extra step improves user control over her account while sharing. On the other hand, the ability to complete the required action remotely for the second factor contradicts with security. In other words, there is an inherent trade-off between real-time phishing resistance and ease of account sharing. 

Security decision makers should take into account the implications of FIDO2 passwordless authentication on delegated access problem and consider deployment of more sophisticated delegation solutions \cite{crampton} in case both access delegation and FIDO2 passwordless authentication are indispensable.  
\\

\fbox{\begin{minipage}{24em}
REASON-4: You cannot easily share account and delegate access with FIDO2 passwordless authentication.
\end{minipage}}

\section{Daily Use Concerns}

Although we do not agree, the issues we discussed up to this point can be considered by some as corner cases which are not as important as everyday use. However, earlier comparative user studies on FIDO2 passwordless authentication showed that there are also user concerns in daily use that impede users’ willingness to abandon passwords. 

As mentioned, users develop many coping strategies for their password protected accounts. While some of these are advised against (writing on a Post-it note), yet some others like password managers are even part of best practices since they could improve security (with the strong password generation feature). For most use cases, the burden of remembering passwords is already not needed. Even when we assume passwords are associated with cognitive load, passwordless authentication is not free, it replaces cognitive effort with physical effort. We consult to recent earlier work to have an understanding about user perceptions about this change.


In a lab study on passwordless authentication using a USB security key (YubiKey), 39\% of participants in the test group criticized the need to carry a device to be able to authenticate. They found it annoying because it is not possible to access their accounts if the key is not present, which definitely hampers with spontaneous and ad hoc use \cite{lyastani}.

In another user study with Yubikeys, after completing three tasks (setup and login for Google, Facebook and Windows 10), participants evaluated the YubiKey using the System Usability Scale (SUS) and gave a mean score of 49.7, which is considered not acceptable (F grade) \cite{reynolds}.

Farke et al. reported on a four-week evaluation period in which they accompanied the deployment of FIDO2 in a company \cite{farke}. They observed that the auto-fill procedure in which there is no physical interaction with additional hardware requires fewer steps than FIDO2 passwordless authentication. As result, it is a significantly faster authentication procedure. Login with a security key takes even longer than logins with manual password entries.

In one of the very few user studies investigating smartphones as FIDO2 authenticators, the authors conducted a between-subjects (N=97) study and found that participants who use passwordless authentication were less likely to authenticate successfully than participants who use passwords \cite{owens}. Passwordless authentication also took significantly longer. Participants in passwordless group had lower agreement that authentication was easy. Passwordless authentication received an “OK” rating in the initial survey and a “Good” rating in the exit survey whereas passwords received an “Excellent” rating in both surveys \cite{owens}. \\  

\fbox{\begin{minipage}{24em}
REASON-5: It is hard if not impossible to beat the convenience of password managers autofilling login forms.  
\end{minipage}}

\section{Other Concerns}

In this section, we overview other hindrances that reduce the chances of FIDO2 getting accepted as a password replacement.

\textbf{Misconceptions about biometric data.} Lassak et al. conducted an empirical study to understand user misconceptions about where the biometric data is stored in FIDO2 protocols \cite{lassak}. They reported that 67\% of participants incorrectly thought their biometrics were sent to the website, creating security concerns. They found that carefully crafted notifications improved perceptions and partially addressed misconceptions, however misconceptions about where the biometric data is stored partially persisted and discourage adoption.

\textbf{Interoperability issues and Switching to a new phone.} Support for FIDO2 standards from big tech companies is a double edge sword. On one side, this support is definitely needed; a solution which only works in a certain browser brand is certainly undesirable. On the other side, \textit{"Historically, it's bad news when big corporations get involved with standardizing anything security related. It generally means it's going to be complicated, convoluted, and hopelessly tied to proprietary stuff you can only get from these big corporations. Interoperability is almost guaranteed to not be a thing} \cite{blog}". 

To illustrate the potential interoperability issues, consider that you want to switch to a new mobile phone (from Android to iPhone or vice versa). First of all, you could do it only if you use so-called multi-device credentials where your FIDO2 private keys are no longer unexportable \cite{fidofaqs, multidevice}. As you see, things already start to be complicated even when brand of your new phone is not changed. 

Suppose you accept the risk coming with this type of credentials. The next issue is the actual transfer of credentials. According to FIDO Alliance's FAQ page, caBLE (cloud-assisted Bluetooth Low Energy) protocol which presumably works across different platforms is used for this purpose. This protocol was developed by Google in 2019 and announced to be also supported by Apple \cite{news2} \footnote{Apple has currently another solution called Passkey \cite{passkey} for synchronizing across devices.}. We could not find any information when this support will be available from Apple official FIDO2 announcement \cite{news}. But we know that Google’s caBLE extension has no public specification \cite{putz} and implemented only in the Chrome web browser as of July 2021.

How about a use case in which a company decides to use its own on-prem servers for the storage of FIDO2 multi-device credentials of its workforce instead of trusting the cloud of Google and Apple? The FIDO2 standards define an extension interface for these purposes and Google developed caBLE protocol as an extension. On the other hand, developing new FIDO2 extensions is currently limited to developers of web browsers or FIDO Alliance members. Third-party developers have encountered many difficulties while implementing and testing their custom FIDO2
extensions \cite{putz}.

\textbf{Aged Population and Other Demographic Factors.} Overcoming adoption barriers we listed for passwordless authentication could be even more difficult for special groups such as older adults. Das et al. reported that participants older than 60 years experienced more difficulty while registering their security key to an email and social media account and it took an average of 52 minutes to complete it whereas with younger adults the same task took around 10-15 minutes \cite{das}.

\textbf{Clever Social Engineering Attacks.} In our opinion, the ability of FIDO2 to safeguard users against advanced threats should be taken with a grain of salt. FIDO2 is not a silver bullet and security experts already know that there is no way to eliminate the weakest link in the security chain, that is the human with only technical countermeasures. We have already mentioned one attack type called Downgrade Attack which undermines the added security of FIDO if weaker 2FA alternatives are supported (for backup) \cite{ulqinaku}. Below, we sketch how a social engineering attack, which does not depend on a weaker alternative, could still be carried out if additional authenticator keys could be defined for account recovery or for other reasons as suggested by FIDO itself \cite{recovery}.

We call our novel attack as \textit{targeted social engineering with a security key gift}. Suppose that due to increasing risks of traditional 2FA methods, a bank decides to replace it with FIDO2 passwordless authentication. Cost factors and the customers' previous experience play a role and smart phones are chosen to be used as primary authenticators. But, with a small fee, an additional USB security key activated with a PIN is made available for paying customers. In our story plot, we imagine a talented attacker targeting a rich customer and who has already obtained his postal address and cell phone number. He sends his victim a small package which includes a security key and a small note saying that it is a gift from the bank for its loyal customers. After making sure that his gift is on target, he calls the victim pretending to be from bank helpdesk. He is so kind to explain the victim how to setup his additional authenticator and he even helps him to determine a strong PIN by suggesting one. After some period of time, the attacker sends the victim a second gift and calls him again requesting him to return back the old one to a mailbox address because of a malfunction or some other technical reason. Bingo!

The attack briefly sketched above is indeed not straightforward. But that does not mean it is not applicable. Daily, we read many types of probably more sophisticated scams (not directly targeting authentication process) in the news. It is worth the effort if the reward is high enough. Our intent here is to illustrate why no technical countermeasure including FIDO2 is secure against attacks exploiting human errors.

\textbf{Applications other than Web.} Our discussion in this paper is intentionally restricted to the problem of user authentication on the web. There is a wide range of applications other than web requiring user authentication. In some of which, passwordless authentication could be a nice and already implemented choice (e.g., Windows Hello), yet in some other legacy systems the support is limited if there is any, as a result, password replacement could be even more challenging.

\section{Conclusion and Future Work Suggestions}

In this position paper, we draw on our collective industry expertise and recent research results to provide main reasons why we think FIDO2 will not be the \textit{kingslayer} of passwords \cite{lyastani}. Without compromising our position, we also expect FIDO2 to be the most favorite solution for passwordless authentication if adopted due to the support from big tech companies.

Despite the risk of being terribly wrong, here is our prediction on the future of passwords and user authentication in two parts:

1) For customer identity, we will see risk based authentication \cite{wiefling} solutions and the use of machine learning techniques, more. After the risk profile associated with a user session is computed, passwords will suffice for low risk profiles. Higher risk profiles lead to stronger options, which include FIDO2. FIDO2 may also be adopted more for particularly vulnerable user accounts and in advanced protection programs such as the one introduced by Google \cite{advancedprogram}.

2) For workforce identity, SSO solutions will become more popular. Being a critical front door to all other applications, SSO logins will require strong passwords and implement FIDO U2F at first. Only some of these may choose to kill passwords and switch to FIDO2 passwordless authentication by time. Risk based authentication techniques will play a larger role for workforce identity as well.

We finish our paper by listing our future work suggestions on (passwordless) authentication:

1) We assert that a major issue contributing to the user reluctance to switch to FIDO2 will be inconsistent and ambiguous setup interfaces and terminology \cite{ciolino}. We want to point out that although the same authenticator could be used for all web sites supporting FIDO2, each web site requires a separate setup. A standard process and consistent user interface for FIDO2 setup is not currently available. We do not know if standards could help here and before introducing any standard, we need more user research to discover the least exhausting setup process for FIDO2 passwordless authentication.   

2) In our view, rather than claiming advantages with respect to usability, persuading users to adopt protective behaviour should play a large role for promoting adoption of password alternatives. In a large scale study, Golla et al showed that user interface design patterns effective in other fields can also increase 2FA adoption \cite{golla}. Similar studies on design and messaging strategies could be carried out for FIDO2 passwordless authentication.

3) Users have legitimate concerns for reaching their accounts when their authenticators become permanently or temporarily unavailable. There are many account recovery options for FIDO2 passwordless authentication, but none of them has gone through a user study (Kunke et al provided a heuristic evaluation for 12 different options, which is useful as a starting point \cite{kunke}). A related research problem is investigation of user perception about multi-device credentials relaxing the security principle that private keys must never leave the authenticator device \cite{kunke}. There may be even misconceptions among both users and developers due to the fact that earlier WebAuthn specification does not permit transferring the private key across authenticators \cite{lassak}.

4) We previously claimed that resisting real-time phishing attacks and keeping the ability for account sharing are at odds with each other. Up to our best knowledge, the trade-off between these two system parameters is not previously touched. Somebody could prove we are wrong by introducing an innovative solution for this problem.

5) As mentioned, since FIDO2 is a very recent standard, there is only a small number of studies directly aiming at evaluating usability of passwordless authentication for end users. Up to our best knowledge, there is not any user study yet about the FIDO2 APIs with respect to their developer usability. We suspect that as being notoriously complex, these APIs are prone to usage errors. In addition, evaluation of available web tools (e.g., WebDevAuthn \cite{webtool} for analyzing FIDO2 messages) could help us to understand their practical utility for developers in debugging and testing.

\end{document}